\documentclass[aip,jap,graphicx,reprint]{revtex4-1}
\usepackage{units}
\usepackage{graphicx}
\usepackage{color}

\draft 

\begin{document}


\title{Low temperature investigations and surface treatments of colloidal narrowband fluorescent nanodiamonds} 



\author{E. Neu}
\affiliation{Universit\"at des Saarlandes, Fachrichtung 7.2, 66123 Saarbr\"ucken, Germany}
\affiliation{Universit\"at Basel, Departement Physik, 4056 Basel, Switzerland}

\author{F. Guldner}
\author{C. Arend}
\affiliation{Universit\"at des Saarlandes, Fachrichtung 7.2, 66123 Saarbr\"ucken, Germany}
\author{Y. Liang}
\affiliation{Universit\"at W\"urzburg, Institut f\"ur Organische Chemie, 97074 W\"urzburg, Germany }
\author{S. Ghodbane}

\author{H. Sternschulte}
\altaffiliation{present address: Fakult\"at f\"ur Physik, Technische Universit\"at M\"unchen, James-Franck-Strasse 1, 85748 Garching, Germany }
\author{\mbox{D. Steinm\"uller-Nethl}}
\affiliation{rho-BeSt Coating Hartstoffbeschichtungs GmbH, 6020 Innsbruck, Austria}

\author{\mbox{A. Krueger}}
 \email{anke.krueger@uni-wuerzburg.de}
\affiliation{Universit\"at W\"urzburg, Institut f\"ur Organische Chemie, 97074 W\"urzburg, Germany }
\author{C. Becher}
 \email{christoph.becher@physik.uni-saarland.de}
\affiliation{Universit\"at des Saarlandes, Fachrichtung 7.2, 66123 Saarbr\"ucken, Germany}


\date{\today}

\begin{abstract}
We report fluorescence investigations and Raman spectroscopy on colloidal nanodiamonds (NDs) obtained via bead assisted sonic disintegration (BASD) of a polycrystalline chemical vapor deposition film. The BASD NDs contain \emph{in situ} created silicon vacancy (SiV) centers. Whereas many NDs exhibit emission from SiV ensembles, we also identify NDs featuring predominant emission from a single bright SiV center. We demonstrate oxidation of the NDs in air as a tool to optimize the crystalline quality of the NDs via removing damaged regions resulting in a reduced ensemble linewidth as well as single photon emission with increased purity. We furthermore investigate the temperature dependent zero-phonon-line fine-structure of a bright single SiV center as well as the polarization properties of its emission and absorption.
\end{abstract}

\pacs{}

\maketitle 

\section{Introduction}
Recently, narrowband fluorescent nanodiamonds (NDs) harnessing the photoluminescence of silicon vacancy (SiV) centers have attracted research interest due to their promising applications as single photon sources as well as fluorescence labels for in vivo imaging.\cite{Neu2011a} For the application as single photon sources, SiV centers stand out owing to high brightness, narrow bandwidth single photon emission: Single photon rates up to 6 Mcps have been reported. Simultaneously, the fluorescence predominantly concentrates in the narrow purely electronic transition, i.e., the zero-phonon-line (ZPL) with down to 0.7 nm width at room temperature.\cite{Neu2011, Neu2012a} For the application as fluorescence labels, SiV centers are promising as they enable optical excitation using red laser light (e.g., at 671 nm) combined with fluorescence in the near-infrared spectral range at approx.\ 738 nm.\cite{Neu2011a} Both properties aid to minimize absorption in biological tissue as well as tissue autofluorescence;\cite{Weissleder2003} additionally, the narrow bandwidth enables efficient spectral filtering of fluorescence and background signal. In Ref.\ \onlinecite{Neu2011a}, we reported the production of an aqueous, colloidal solution of SiV containing NDs from polycrystalline chemical vapor deposition (CVD) diamond films using the bead assisted sonic disintegration (BASD) method (see Refs.\ \onlinecite{Ozawa2007,Liang2009}). These SiV containing NDs in solution are promising for applications as fluorescence labels as well as single photon sources. However, some essential characteristics of the SiV fluorescence in BASD NDs have not been investigated to date.

In the context of single photon sources, the feasible positioning of NDs in solution by nano-manipulation techniques is advantageous: It enables the coupling to photonic nano-structures, e.g., dielectric resonator structures to build enhanced single photon sources.\cite{Benson2011} To reach efficient coupling of a defect center to a resonator mode, spectral overlap of the resonator mode and the color center's emission line has to be ensured.\cite{Riedrichmoeller2011} Cooling of the color center generally reduces the emission linewidth and thus potentially enhances the spectral overlap of an emission line with a narrow resonator mode of a high Q cavity. However, the cooling also induces a spectral shift of the ZPL.\cite{Neu2011} Thus, knowledge of the ZPL shift as well as the temperature dependent linewidth is crucial to ensure spectral overlap at the targeted temperature when coupling color centers to resonator structures. For SiV centers in BASD NDs, neither temperature dependent linewidths nor lineshifts have been investigated to date.

Second, the polarization of single photons emitted from a color center is crucial, e.g., in quantum key distribution applications (Ref.\ \onlinecite{Bennett1984}) or for the frequency conversion of single photons.\cite{Zaske2011} At cryogenic temperature, the ZPL of the SiV center splits into 4 fine-structure components.\cite{Sternschulte1994} However, the polarization properties of these line components have not been reported in the literature in contrast to the room temperature polarization properties.\cite{Neu2011b}

Third, for single photon sources as well as fluorescence labels, spectrally and temporally stable fluorescence emission is essential. Therefore, stabilizing the desired charge state of the color center, corresponding to defined luminescence properties, is crucial. For nitrogen vacancy color centers in NDs, it has been shown that the charge state critically depends on the surface termination of the NDs.\cite{Rondin2010, Petrakova2012} Nevertheless, no data on the dependence of SiV luminescence in NDs on the surface termination is reported in the literature. Using heating in air (oxidation), amorphous carbon and graphitic shells can be preferentially removed before the $sp^3$ bonded diamond core of the ND is reduced.\cite{Gaebel2012} At 400--430 $^\circ$C, $sp^2$ bonded carbon can be etched without loss of diamond even for diamond nanoparticles.\cite{Osswald2006} Simultaneously, the surface is covered with oxygen containing functional groups such as carboxyl and carbonyl groups.\cite{Petrakova2012} At higher temperatures, diamond can be etched away and the NDs can be reduced in size: At 500 $^\circ$C, HPHT NDs with about 60 nm size are etched at a rate of 1 nm/hour.\cite{Gaebel2012} However, the etch rate strongly depends on the crystallographic direction.\cite{Gaebel2012} Furthermore, smaller diamonds get etched or even destroyed at lower temperature (see Ref.\ \onlinecite{Osswald2006} for work on detonation NDs).
In the present work, we use air oxidation to remove damaged regions from the surface and establish a defined surface termination of the BASD NDs. We investigate the effects of the surface treatment via Raman spectroscopy and we observe the reaction of the SiV center luminescence to the same treatments. We furthermore address the crucial topics introduced above: temperature dependent linewidth and position as well as the polarization of the ZPL components.
\section{Investigation of the surface treatment process via Raman spectroscopy \label{sec_Raman}}
\begin{figure}
\centering
\includegraphics{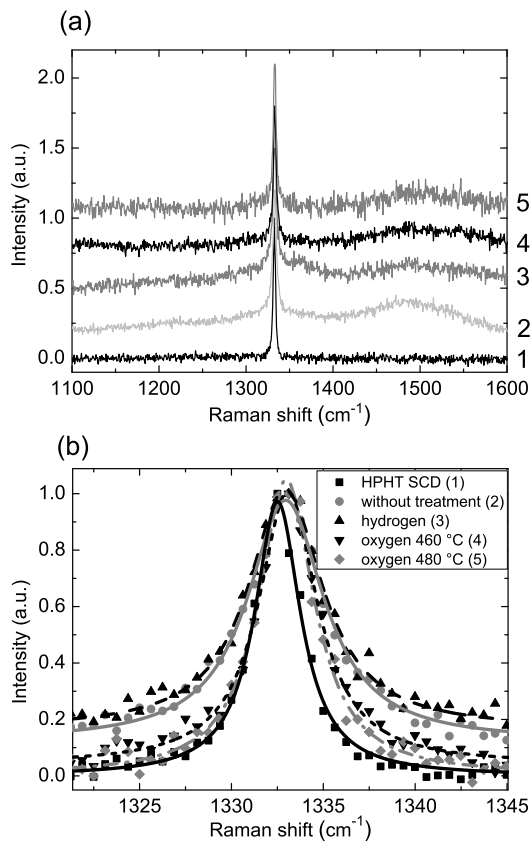}
\caption{(a) Raman spectra of an ensemble of BASD NDs after different treatments: curve 2 without treatment, curve 3 hydrogen treated, curve 4 oxygen 460 $^\circ$C, curve 5 oxygen 480 $^\circ$C.  For comparison, the Raman spectrum of a high pressure high temperature single crystalline diamond is given (curve 1). The spectra have been recorded using 488 nm laser light, corrected for a linearly rising fluorescence background and shifted vertically for clarity. (b) detailed spectrum of the diamond Raman line with Lorentzian fit. Note that the calibration of the Raman Spectrometer has been done on the Raman line of an HPHT single crystal diamond set to 1332.5 cm$^{-1}$.  \label{figureraman}}
\end{figure}

The investigated NDs have been described in detail in Ref.\ \onlinecite{Neu2011a}. They are obtained via the BASD method from a polycrystalline CVD diamond film grown on a Si substrate. SiV centers are created \textit{in situ} as a result of the etching of the Si substrate and subsequent incorporation of Si into the growing diamond. The NDs have a size of 70--80 nm. For the following measurements, we use samples where the colloidal solution has been dropcast and dried on an Ir multilayer substrate.\cite{Gsell2004} A Raman spectrum of an ensemble of untreated NDs is depicted in curve 2 in Fig.\ \ref{figureraman}(a). Raman spectroscopy confirms a high crystalline quality of the NDs: we find a Raman line at 1333.00 $\pm$ 0.07 cm$^{-1}$ with a width of 4.9 $\pm$ 0.4 cm$^{-1}$. The error is obtained as the standard deviation of the width and position measured on different spots on the dropcast sample. To obtain the width, a Lorentzian fit has been used. The observation of a Raman line shifted only by 0.5 cm$^{-1}$ to higher frequency compared to the case of single crystal diamond witnesses a low absolute (compressive) stress in the diamonds (see, e.g., Ref.\ \onlinecite{Bergman1995}) whereas the broadening of the line indicates the stress distribution.\cite{Kirillov1994} However, the Raman spectrum also reveals a signature of $sp^2$ bonded carbon at around 1500 cm$^{-1}$.\cite{Zaitsev2001} We note that a very similar non-diamond Raman signature has also been present in the CVD diamond film used as a starting material for ND production.\cite{Neu2011a} The $sp^2$ bonded carbon is supposed to reside mainly on or below the surfaces of the diamonds partly as a result of surface damage during the milling procedure or as a residual of $sp^2$ bonded carbon from the grain boundaries of the CVD film. As a consequence of the milling process being carried out in water, the NDs most probably initially reveal a termination with hydroxyl functional groups.\cite{Liang2009} However, acid cleaning has been subsequently employed to remove contaminants from the milling process possibly partially oxidizing the surfaces.

We now subject the sample to two oxidation steps: In a first oxidation step, the sample is heated to 460 $^\circ$C in air and kept at that temperature for 90 mins [Raman spectrum: curve 4 in Fig.\ \ref{figureraman}(a)]. In a second oxidation step, we keep the sample at 480 $^\circ$C for 40 mins [Raman spectrum: curve 5 in Fig.\ \ref{figureraman}(a)]. After the heating steps, the sample is cooled down inside the oven. We additionally test the effect of a hydrogen plasma (hot filament plasma, 33 mins, pressure $\approx$12 mbar, distance to filament 5-6 mm, filament current 16 A, temperature (sample holder) 500 $^\circ$C) on untreated BASD NDs [Raman spectrum: curve 3 in Fig.\ \ref{figureraman}(a)]. All surface treatments, especially the heating in air to 480 $^\circ$C reduce the signal related to $sp^2$ carbon which is in accordance with the assumption that the $sp^2$ carbon phase is located at the surfaces of the NDs. Furthermore, the oxygen surface treatments narrow the diamond Raman line in contrast to the hydrogen treatment: After the first oxidation step, we find a width of 4.0 $\pm$ 0.2 cm$^{-1}$; following the second step it further reduces to 3.7 $\pm$ 0.2 cm$^{-1}$. As mentioned before, the stress distribution in the diamond material can induce a broadening of the Raman line.\cite{Kirillov1994} Furthermore, diamond with a high defect density has a broader Raman line.\cite{Kirillov1994} Thus, the narrowing of the line suggests that the oxygen treatment is capable of removing highly stressed as well as highly defective regions of the milled NDs. The oxidation procedure thus constitutes an enhancement of the crystalline quality of the NDs. We note that further heating in air (500 $^\circ$C, 2 hours), led to a significant reduction of the diamond quantity and no Raman spectra could be obtained after that heating step. As a preliminary remark for the next section, we note that luminescence of SiV centers has been observed on the untreated as well hydrogen treated and oxidized dropcast samples.
\section{Room temperature investigations of SiV ensembles and single SiV center}
To enable the investigation of the fluorescence of individual NDs, we spin coat the aqueous solution onto an Ir multi-layer substrate.\cite{Gsell2004} These substrates provide low background fluorescence for the luminescence investigation of single SiV centers\cite{Neu2011} and highly efficient fluorescence collection.\cite{Neu2012a} Moreover, these substrates withstand the oxidation steps which we carry out. Due to randomly created spatial patterns of the spin coated diamonds, we reliably identify and relocate distinct NDs on the untreated samples as well as after the oxidation steps in a homebuilt confocal microscope setup.  The employed confocal setup uses a microscope objective with a numerical aperture of 0.8 and the internal detection efficiency of the setup is estimated as 25\%. For more details on the experimental setup see, e.g., Ref.\ \onlinecite{Neu2011}.

\begin{figure}
\centering
\includegraphics{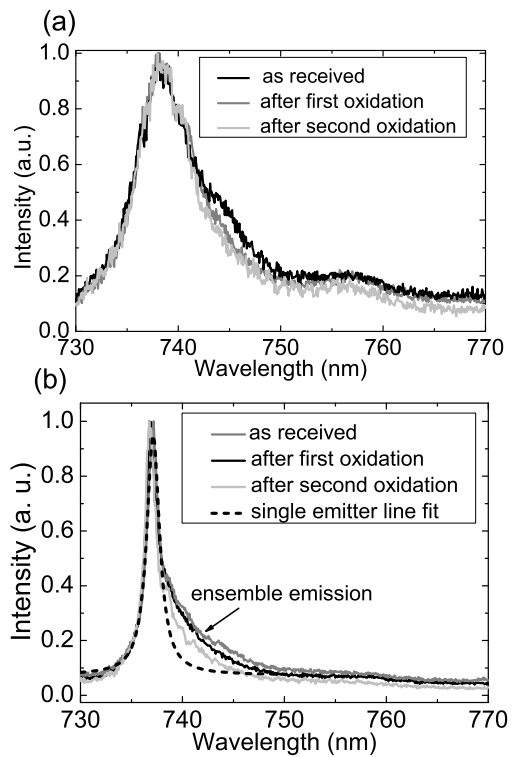}
\caption{(a) Spectrum of SiV ensemble in BASD nanodiamond and changes during oxidation, (b) spectrum of ND with dominant emission from a single SiV center (Fit: dashed line, peak 737.2 nm, width 1.5 nm) with underlying ensemble emission especially visible in the long wavelength edge of the single emitter spectrum. \label{figurespectra}}
\end{figure}
\begin{figure}
\centering
\includegraphics{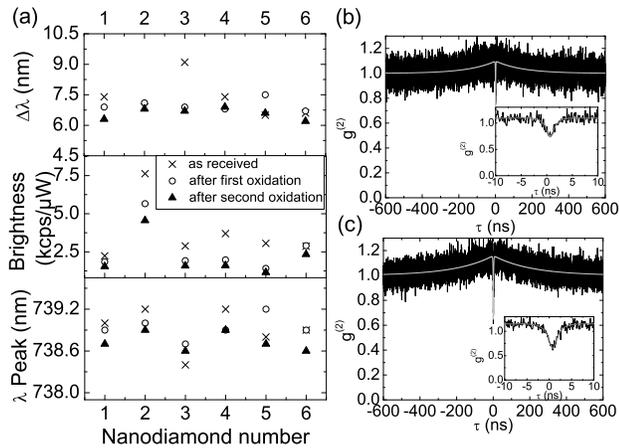}
\caption{(a) Changes in ZPL width $\Delta \lambda$, brightness and ZPL peak wavelength $\lambda$ for NDs containing SiV ensembles. (b) $g^{(2)}$ function of a single SiV center with underlying ensemble emission after first oxidation step, (c) $g^{(2)}$ function of a single SiV center with underlying ensemble emission after second oxidation step. Gray solid lines represent fits to the data using the $g^{(2)}$ function of a three level system convoluted with the instrument response of the measurement setup and corrected for background emission (for details see e.g.\ Ref.\ \onlinecite{Neu2011}). Both $g^{(2)}$ measurements have been performed using 20 $\mu$W of excitation power.  \label{figureoxtrendg2} }
\end{figure}

The untreated NDs mostly contain ensembles of SiV centers: We observe no saturation of the fluorescence and intensity auto-correlation ($g^{(2)}$) measurements do not indicate single photon emission (antibunching). An example of the typical luminescence spectrum of an ensemble of SiV centers is displayed in Fig.\ \ref{figurespectra}(a): The ZPL of the SiV center is clearly visible, here at 739 nm. For most NDs, the ZPL resembles an asymmetric tail towards longer wavelengths, indicating an inhomogeneous broadening of the SiV center ensemble ZPL with a preferential red shift.\cite{Clark1995} Due to the asymmetry as well as the unknown fraction of homogenous and inhomogeneous broadening, we can only estimate the linewidth $\Delta \lambda$  as well as the peak position $\lambda$. We use a Lorentzian fit to the data as a rough estimate. We obtain linewidths ranging from 6.5 to 9.1 nm, with peak wavelengths between 738.8 and 739.2 nm for the investigated NDs in accordance with our previous observations in Ref.\ \onlinecite{Neu2011a}  [see Fig.\ \ref{figureoxtrendg2}(a)]. To estimate the brightness of the NDs, we measure the fluorescence intensity vs. excitation power for each ND. As a measure of the brightness, we employ the slope of the linearly rising intensity vs.\ excitation power curve (in cps/$\mu$W). The results are displayed in Fig.\ \ref{figureoxtrendg2}(a). Using 671 nm excitation, we obtain on average 3900 cps/$\mu$W without any indication of saturation or fluorescence instability up to roughly 0.5 mW. Thus, these NDs containing SiV ensembles are potential candidates for stable, narrowband fluorescence markers.

We now subject the sample to the oxidation steps introduced in Sec.\ \ref{sec_Raman}. The ZPL linewidth $\Delta \lambda$, the brightness as well as the ZPL peak wavelength $\lambda$ for 6 NDs is summarized in Fig.\ \ref{figureoxtrendg2}(a). We observe a general trend, which is also discernible in Fig.\ \ref{figurespectra}(a): The ZPL linewidth is reduced, especially due to a narrowing of the line at the red wavelength edge, thus slightly diminishing the asymmetry of the line. Due to this reduction of the asymmetry, we also obtain a shift of the peak wavelength when fitting a Lorentzian line to the data. The results are summarized in Fig.\ \ref{figureoxtrendg2}(a). The maximum shift of the peak position is 0.3 nm, the typical reduction of the linewidth is 0.5--1 nm.  Furthermore, changes in the sideband spectrum, including vanishing features in the sideband region, have been observed for several NDs. Assuming that the absorption coefficient and the brightness of the SiV centers as well as the collection and detection efficiency remain unchanged, the brightness of a ND before and after oxidation can serve as a measure of the number of SiV centers contained in the ND. We observe a trend to a reduced brightness of the NDs after each oxidation step; in particular, after the second oxidation step, the average brightness is reduced to 2400 cps/$\mu$W. Thus, one might conclude that the number of SiV centers in the NDs is reduced. The ZPL is narrowed during the oxidation steps, as probably predominantly red shifted SiV centers in stressed regions close to the surface of the NDs are removed. The cleaning of the surface, however, does not seem to enhance the brightness of the remaining SiV centers. This observation is in contrast to the observations reported in Ref.\ \onlinecite{Iakoubovskii2000a} were non-diamond carbon especially in the grain boundaries of polycrystalline diamond films has been found to quench SiV luminescence. Thus, one might expect a significant enhancement of the fluorescence upon surface cleaning. We emphasize that as we do not know how many centers have been removed, we cannot conclusively determine the influence on the brightness of single SiV centers, especially close to the surface.

In contrast to the spectra of SiV ensembles discussed above, we also find NDs where a significantly narrowed, bright line dominates over a broader emission. An example of such a spectrum is given in Fig.\ \ref{figurespectra}(b). A Lorentzian fit of the narrow line yields a peak position of 737.2 nm with a width of 1.5 nm [see dashed line in Fig.\ \ref{figurespectra}(b)]. Additionally, a broader emission is present that is especially visible in the long wavelength tail of the spectrum [marked with an arrow in Fig.\ \ref{figurespectra}(b)]. We interpret this spectrum as follows: The narrow bright line originates from the emission of a single, bright SiV center, whereas the underlying broad emission originates from the inhomogeneously  broadened emission of an ensemble of weakly fluorescing SiV centers. The observed linewidth of the single SiV center is in accordance with previous observations for the linewidth of single SiV centers (0.7-2.2 nm, Ref.\ \onlinecite{Neu2011}).  We note that in Ref.\ \onlinecite{Neu2012a} single SiV centers with brightness differing by almost two orders of magnitude have been observed, thus indicating a large spread of single emitter brightness for SiV centers, suggesting why a single emitter can dominate over a whole ensemble of centers. 

To prove this interpretation of the spectrum, we perform $g^{(2)}$ measurements after the first oxidation step. The $g^{(2)}$  measurement shows an antibunching dip and is displayed in Fig 3(b). However, $g^{(2)}(0)=0.75$, thus only imperfect single photon emission is observed and one cannot straightforwardly conclude that a single emitter is present. In the following, we discuss in detail that the measured $g^{(2)}$ function is in accordance with our interpretation of the spectrum consisting of a narrow single emitter line and an underlying ensemble emission. The $g^{(2)}$ function indicates a three level internal population dynamics in accordance with previous observations on single SiV centers.\cite{Neu2011,Neu2012a} We fit the correlation measurements using the $g^{(2)}$ function of a three level system. To take into account the timing jitter of the detection setup, which leads to a deviation from $g^{(2)}(0)=0$, we convolute the $g^{(2)}$  function with the instrument response of our Hanbury Brown Twiss setup (timing jitter 354 ps, Gaussian jitter $\frac{1}{\sqrt{e}}$ width, for details see Ref. \onlinecite{Neu2011}).  Furthermore, we include the underlying ensemble emission as an uncorrelated background following Ref.\ \onlinecite{Beveratos2001}. The assumption of an uncorrelated emission from the ensemble is justified as for NDs containing ensembles of SiV centers, we only found featureless $g^{(2)}$  functions indicating uncorrelated emission.  We use the spectrum in Fig.\ \ref{figurespectra}(b) to estimate the ratio between the emission intensities in the single emitter line as well as in the ensemble. We note that for the $g^{(2)}$   measurements a bandpass filter with a transmission window between 730 and 750 nm has been used, thus photons from the ensemble emission as well as the single emitter contribute to the measurement. We find a probability of 63\% that a detected photon has been emitted into the narrow line and thus stems from the single emitter.  We include this probability into the fit of the $g^{(2)}$  function to account for the background emission and obtain a very good agreement with the measured data [see solid lines Fig.\ \ref{figureoxtrendg2}(b)] thus proving our interpretation that the main contribution to the emission, i.e. the narrow, bright emission line, stems from a single SiV center.

After the second oxidation, a more pronounced antibunching dip with $g^{(2)}(0)=0.6$ is observed [see Fig.\ \ref{figureoxtrendg2}(c)].
Furthermore, following the second oxidation step, we clearly observe a saturation behavior as expected for a single emitter instead of a linearly rising fluorescence. We use
\begin{equation}
      I(P)=I_\infty\frac{P}{P+P_{sat}}+c_{backgr}P
      \label{Satfunc2}
\end{equation}
to fit the fluorescence intensity $I$ in dependence of the excitation power $P.$ Here, $P_{sat}$ is the saturation power, where $I_{\infty}$ is the maximum emission rate of the single emitter and $c_{backgr}$ gives the brightness of uncorrelated background emission, here predominantly due to the underlying SiV ensemble. The fit yields $P_{sat}=111$ $\mu$W, $I_{\infty}=581$ kcps, $c_{backgr}=2000$ cps/$\mu$W.
From this saturation measurement, we estimate the probability that a photon has been emitted by the single, bright SiV center in the $g^{(2)}$ measurement as 69\%. Again, the fitted $g^{(2)}$ function well describes the measured data, proving that the bright, narrow ZPL belongs to a single emitter. Thus, the underlying ensemble emission is reduced by the second oxidation step as also visible from the spectrum displayed in  Fig.\ \ref{figurespectra}(b), where a significant narrowing of the line at the red edge of the ZPL is discernible. Thus, the observation coincides with the findings on the NDs hosting SiV center ensembles. The fact that the narrow single emitter is not destroyed during the oxidation steps hints at an incorporation of this color center in a region of the ND which is not particularly close to the surface. Summarizing, we have successfully employed oxidation in air as a method to reduce the influence of a weakly fluorescing ensemble of SiV centers resulting in an increased single photon emission probability from a bright single SiV center in a BASD ND, thus significantly enhancing its performance as a single photon source.

\section{Low temperature characterization of a single SiV center}
\begin{figure}
\centering
\includegraphics{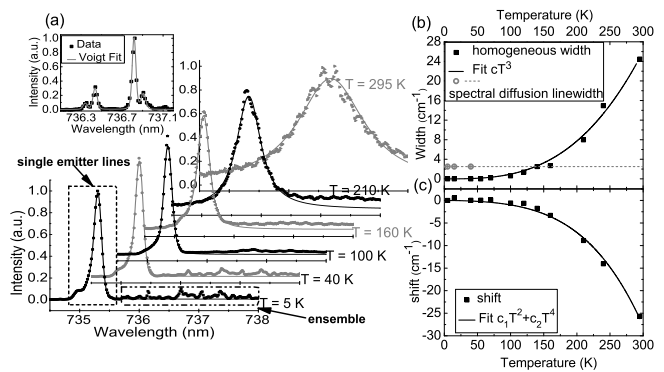}
\caption{(a) Temperature dependent spectra of a BASD ND with dominant emission from a single SiV center, excitation 671 nm. Shifted to longer wavelength from the main line of the single SiV center, the underlying ensemble is clearly visible. The inset shows the ZPL fine-structure of an ensemble of SiV centers in a thin high quality homoepitaxial CVD diamond film at 6.4 K for comparison. (b) homogeneous and inhomogeneous (spectral diffusion) width of the main line of the single SiV center, (c) shift of the main line of the single center.  \label{figtempdep}}
\end{figure}
\begin{figure}[h]
\centering
\includegraphics{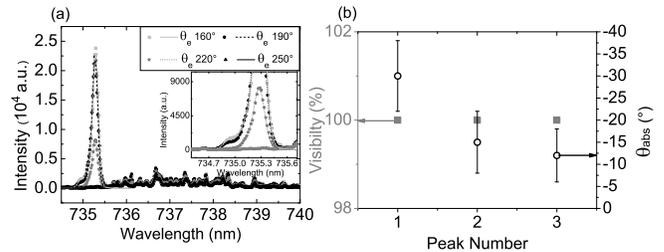}
\caption{Polarized absorption measured on a ND with dominant emission from a single SiV center and underlying ensemble emission (a) fluorescence spectra taken at different angles of incident excitation polarization $\theta_e$ ($T=15$ K, excitation 0.18 mW). Shifted to longer wavelength from the main line of the single SiV center, the underlying ensemble is clearly visible (wavelength range 736-740 nm). (Inset: enlarged detail of single emitter line), (b) summary of visibility and linear polarization direction corresponding to maximum absorption $\theta_{abs}$for the single emitter lines \label{figpolexc}}
\end{figure}
\begin{figure}[h]
\centering
\includegraphics{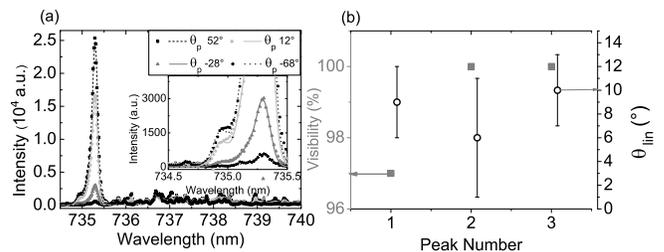}
\caption{Fluorescence polarization measured on a ND with dominant emission from a single SiV center and underlying ensemble emission (a) fluorescence spectra taken at different angles of polarization analyzer $\theta_p$ ($T=15$ K, excitation 0.18 mW). Shifted to longer wavelength from the main line of the single SiV center, the underlying ensemble is clearly visible (wavelength range 736-740 nm). The excitation polarization is fixed to the optimal value. (Inset: enlarged detail of single emitter line), (b) summary of visibility and linear polarization direction $\theta_{lin}$ for the single emitter lines  \label{figpolem}}
\end{figure}

 At room temperature, the ZPL of the SiV center is broadened, possibly masking an underlying fine-structure of the line. To gain further information on the line structure as well as on line shifting and broadening, we thus investigate the luminescence at temperatures down to liquid helium temperature. For the low temperature investigations, we select the previously discussed ND with dominant emission from a single SiV center and weak underlying ensemble luminescence. At temperatures below 100 K, a fine-structure of the ZPL of the single emitter evolves as displayed in Fig.\ \ref{figtempdep}(a). Note that the low temperature characterization has been performed following the first oxidation step. For the weakly fluorescing ensemble accompanying the single SiV center, a complicated fine-structure evolves, red shifted to the single emitter line [see e.g. the spectrum at 5 K in Fig.\ \ref{figtempdep}(a), wavelength range 736-738 nm, or Figs.\ \ref{figpolexc}(a) and \ref{figpolem}(a)].  We interpret this complicated fine structure in terms of overlapping ZPL fine-structure components of single emitters or small sub-ensembles under different stress conditions. The work of Clark et al.\ \cite{Clark1995} and Sternschulte et al.\ \cite{Sternschulte1994} as well as our previous work\cite{Neu2011a} found a four line fine-structure for the ZPL of an ensemble of SiV centers in high quality single crystalline diamond. The four lines arising from a split excited and ground state, arrange in two doublets as discernible from the spectrum in the inset of Fig.\ \ref{figtempdep}(a) measured on a thin high quality homoepitaxial CVD diamond film. The smaller spacing has been experimentally found to represent the ground state splitting of approx.\ 50 GHz and the larger spacing corresponds to the excited state splitting of 240 GHz. The emitter observed here shows a significantly altered and poorly resolved fine-structure fitted best using a dominant main line with two weaker peaks. To explain the observed fine structure one has to remark that the ZPL fine-structure pattern has been observed to change significantly under stress:\cite{Neu2012b,Sternschulte1994} The variations include changes in the relative intensity possibly leading to the disappearance of line components as well as changes in the level splittings and positions. Changes in the fine-structure spectra have been observed for an SiV ensemble under defined stress in [100] direction in Ref.\  \onlinecite{Sternschulte1994} as well as for single centers in CVD grown NDs (unknown stress state) in Ref.\ \onlinecite{Neu2012b}. Thus, also the strongly altered fine-structure of the single emitter in Fig.\ \ref{figtempdep}(a) might originate from an SiV center experiencing an unknown stress state. It should be noted that it is also possible that weak fine-structure components remained unidentified due to overlap with the underlying ensemble.

To evaluate the temperature dependent linewidth as well as the peak position of the ZPL fine-structure components and to separate Gaussian (inhomogeneous) and Lorentzian (homogeneous) contributions to the linewidth, we fit the measured data with three Voigt profiles (for details of the procedure see Ref.\ \onlinecite{Neu2012b}). Fig.\ \ref{figtempdep}(b) summarizes the results for the main peak as only this peak is clearly visible over a broad temperature range [see Fig.\ \ref{figtempdep}(a)]. The fits reveal that at low temperature Gaussian line broadening dominates; the lines are broadened by spectral diffusion. The term 'spectral diffusion' denotes line broadening induced by the Stark shift due to temporally fluctuating charges in the environment of the color center.  We find a spectral diffusion linewidth of 76(1) GHz (2.5 cm$^{-1}$). The spectral diffusion contribution is constant below approx.\ 50 K. In the following, we will discuss a possible source of the fluctuating charges leading to spectral diffusion.  Ref. \onlinecite{Wolters2013}  finds that the main source of spectral diffusion are charges created as a result of the laser excitation of the diamond.  Under this assumption, spectral diffusion is expected to be temperature independent which agrees well with our observation. A major source for laser excited carriers is substitutional nitrogen which is a prevalent impurity in diamond and can be ionized by laser light with a photon energy exceeding 1.7 eV (wavelength 729 nm).\cite{Wolters2013,Tamarat2006}  As a result, the low temperature linewidth for nitrogen vacancy centers has been found to depend strongly on the concentration of substitutional nitrogen in the host diamond: spectral diffusion linewidths for nitrogen vacancy centers in nitrogen rich type Ib diamond can amount to 90 GHz. \cite{Tamarat2006} The spectral diffusion linewidth observed here is thus comparable. In Ref. \onlinecite{Neu2012b}, spectral diffusion linewidths for single SiV centers in the range of 25-160 GHz have been reported. We note that we cannot estimate the nitrogen content of our diamond film, however, no measures have been taken to avoid nitrogen incorporation. 

The homogeneous linewidth due to phonon dephasing\cite{Davies1981} increases with temperature $T$ and is best described using a $\Delta \lambda= c T^3$ law, in accordance with observations in CVD grown NDs.\cite{Neu2012b} The constant $c$ is fitted to be $9.6\pm0.3\times10^{-7}$ K$^{-3}$cm$^{-1}$. A $T^3$ law for the homogeneous line broadening has been attributed to phonon dephasing in a defect rich crystal.\cite{Hizhnyakov1999}

Simultaneously, the main fine-structure component is red shifted with increasing temperature as shown in Fig.\ \ref{figtempdep}(c). From liquid helium temperature to room temperature, we find an overall red shift of 25.7 cm$^{-1}$ in accordance with previous observations of 26 cm$^{-1}$.\cite{Gorokhovsky1995} In Ref.\ \onlinecite{Hizhnyakov2002}, a $c_1T^2+c_2T^4$ dependence of the line shift is deduced as a result of interaction with lattice and local phonon modes and experimentally verified for nitrogen vacancy centers. We adapt this model and yield a very good agreement with the measured data as discernible from Fig.\ \ref{figtempdep}(c). The fitted constants are $c_1= -7\pm2 \times10^{-5}$ K$^{-2}$cm$^{-1}$, $c_2= -2.6\pm0.3 \times10^{-9}$ K$^{-4}$cm$^{-1}$.


We now analyze the polarization properties of the investigated single SiV center as well as the underlying ensemble emission. First, we determine the polarized absorption for the previously investigated ND with dominant emission from a single SiV center via rotating the linear polarization of the excitation laser $\theta_e$ and recording spectra for the different polarizations. The resulting spectra are displayed in Fig.\ \ref{figpolexc}(a). Via fitting the spectrum of the single SiV center, we obtain the areas $A$ of the fine-structure peaks in dependence of $\theta_e$. Fitting a sinusoidal function
\begin{equation}
A(\theta_e)=A_0+A_a\sin^2(\theta_e-\theta_{abs})
\end{equation}
yields the visibility $V$
\begin{equation}
 V=\frac{A_a}{A_a+2A_0},
\end{equation}
for each line as well as the direction of maximum absorption $\theta_{abs}$. The results are shown in Fig.\ \ref{figpolexc}(b). The high visibility close to 100\% indicates that the color center preferentially absorbs linearly polarized light. Furthermore, the direction of maximum absorption $\theta_{abs}$ is very similar for all line components. We thus tentatively conclude that the fine-structure components are excited via the same off-resonant pumping transition. We note that, due to the different orientations of the centers, the polarization reaction of the ensemble (see Fig.\ \ref{figpolexc}(a), wavelength range 736-740 nm) mostly averages out, we thus do not analyze the polarization behavior of the ensemble in detail.

Second, we analyze the polarization properties of the emitted fluorescence via rotating a linear polarization analyzer in the detection path of the confocal setup. $\theta_e$ is fixed to the optimum direction. The resulting spectra are given in Fig.\ \ref{figpolem}(a). We evaluate the spectra analogous to the polarized absorption spectra. We obtain again high visibility for all single emitter lines, thus the emission of the color center is linearly polarized on all fine-structure lines whereas the ensemble emission is only weakly polarized. The direction of the linear polarization of the single emitter lines is equal within 4$^\circ$.
Summarizing, the observed single SiV center shows fully polarized absorption and emission which is advantageous for the application as single photon source as well as fluorescence marker: Via choosing the matching excitation polarization, an efficient excitation is possible suppressing background fluorescence. The emission of fully polarized single photons is highly desirable for applications in quantum key distribution and frequency conversion of single photons. The observations are in accordance with previous findings at room temperature that also identified preferential absorption of linearly polarized light as well as polarized emission.\cite{Brown1995,Neu2011b} To understand the origin of the polarized absorption and emission further investigations using more emitters as well as identification of the transitions are necessary.
\section{Conclusion}
Summarizing, we have successfully employed oxidation in air at up to 480$^\circ$C as a post-fabrication treatment to enhance the crystalline quality of BASD NDs: We find a narrowing of the diamond Raman line from 4.9 to 3.7 cm$^{-1}$ whereas the Raman signal due to non-diamond carbon is significantly reduced. Simultaneously, the linewidth of the SiV ensemble fluorescence in individual NDs is narrowed by up to 1 nm. For a single bright SiV center with underlying ensemble fluorescence, the oxidation leads to an enhancement of the single photon emission probability. Using temperature dependent luminescence measurements (5--295 K), we investigate the fine-structure of the single center's ZPL. At low temperature, the line is broadened by spectral diffusion (76 GHz), whereas homogeneous broadening proportional to $T^3$ dominates at higher temperature. Simultaneously, the line exhibits an overall temperature dependent shift of 25.7 cm$^{-1}$. The center's fine-structure components show a common linear polarization in emission and are excited via the absorption of linearly polarized light. As an outlook, the production of NDs from highly Si doped diamond films could be harnessed to produce brighter NDs as fluorescence labels, whereas the use of high purity material will be advantageous for single photon applications.

%
%

%

\begin{acknowledgments}
We thank M. Fischer, S. Gsell and M. Schreck for supplying the Ir substrates, P. Elens and C. Hepp for help with the fluorescence measurements and A. Fuchs and H. Schmitt for performing the hydrogen treatment of the diamonds. We acknowledge funding by the BMBF (EPHQUAM 01BL0903) and the Deutsche Forschungsgemeinschaft (FOR 1493).
\end{acknowledgments}

%

\end{document}